# PhD bridge programmes as engines for access, diversity and inclusion

Alexander L. Rudolph [1]*, Kelly Holley-Bockelmann[2] and Julie Posselt[3]

**The lack of diversity in physics and astronomy PhD programmes is well known but has not improved despite decades of efforts. PhD bridge programmes provide an asset-based model to help overcome the societal and disciplinary obstacles to improving access and inclusion for students from underrepresented groups and are beginning to show some success. We describe several well-known PhD bridge programmes in the United States and discuss lessons learned from their experiences. Many of these lessons can be extended more broadly to physics and astronomy PhD programmes to increase access, diversity and inclusion.**

Transitions are well-documented times of vulnerability for students navigating the United States' educational system. The impacts of disparate access to high-quality schools accumulate over a student's education, making the transition into graduate education a point at which inequities are especially likely to manifest (and indeed they do). A substantially smaller share of Black/African American, Latinx/Hispanic and Indigenous/Native American (typically identified as underrepresented minority or URM) bachelor's degree recipients in the United States go on to graduate education than their White- and Asian-identifying counterparts[1]. One major reason for this disparity is that a higher proportion of URM students are first generation: in 2011–2012, 42% of Black students and 48% of Latinx students were first-generation, compared with 28% of White students[2].

These disparities are especially profound in science, technology, engineering and mathematics (STEM) disciplines and they are reinforced over time by racialized barriers that are specific to the transition to graduate education. Typical doctoral admissions criteria and processes in the United States limit access to graduate school for students from groups that are already underrepresented[3]; for example, PhD programmes make initial judgments of admissibility on academic metrics and, in doing so, are also likely to rule out students of colour. Black and Latinx students are substantially more likely to attend public than private undergraduate institutions where the mean grades awarded are about one-third of a letter lower than in private colleges. Faculty members favour PhD applicants with elite institutional affiliations, but such institutions' undergraduate admissions processes can be as exclusionary as those at the graduate level[4]. At both levels, standardized admission test scores are a central feature in operational definitions of merit; however, scores vary considerably by gender, race and social class[5,6]. Despite the growing test-optional movement in the US (cheekily dubbed #GRExit at the graduate level) — through which more and more colleges, universities, graduate programmes and graduate schools are eliminating standardized admission test score requirements[7] — and evidence that graduate record examination (GRE) scores are unreliable predictors of PhD completion[8,9], these scores still play a prominent role in most STEM PhD admissions processes[10]. Entrenched assumptions about who deserves admission thus constitute an invisible, racialized barrier in the transition to graduate education.

Financial barriers also impede access to graduate education. With rising college costs, many undergraduate students at large public universities struggle to pay rent and buy food, often working 20–30 hours per week or more[11]. The rigours of such a schedule limit the time available for studying and participation in the extended research experiences that PhD programmes increasingly seek in prospective students[3]. Rising costs also come with growing debt burdens that deter continuation to graduate school[12].

Furthermore, faculty members, institutions and society alike create psychological barriers by sending students of colour mixed messages about their worth and belonging in science[13]. Professors respond less often and more slowly to prospective applicants whose names suggest that they are women and/or from racially minoritized groups[14]. The absence of role models who share one's identity as a source of inspiration or advice compounds these threats[15].

No single intervention can address this system of admissions, financial and psychological barriers; however, one type of programmatic intervention — the PhD bridge programme — has spread in the past 20 years. When implemented with care, it seems to chip away at the myths that pervade faculty thinking about who can be successful and empower talented students to see themselves in academia and take steps in that direction.

## A review of existing programmes

Before formal PhD bridge programmes existed, students who were denied admission to PhD programmes in physics and astronomy as undergraduates had the option to pursue a master's degree as a stepping stone to a PhD. The 23 campuses of the California State University (CSU) system house ten master's programmes in physics or astronomy that play this role. Furthermore, historically Black colleges and universities (colleges and universities founded with the explicit goal of educating Black students) have played such a role for many years; in fact, URM students are 50% more likely to obtain a master's degree on their way to a PhD than their non-minority counterparts[16]. The additional coursework of a master's programme enables access to courses and knowledge that may have been unavailable to them as undergraduates and, for some students, their master's research is their first opportunity to gain research experience that PhD programmes prize highly. The CSU and some other physics and astronomy master's programmes time their admissions to allow students who have been denied admission to a PhD programme to apply during the spring of their graduating year, thereby permitting them to matriculate that fall; in other cases, early master's admissions deadlines force students to wait an additional year before matriculating to a master's programme. In either case, the need to take the extra step of attending a master's programme can often delay admission to a PhD programme by an additional year or more, in addition to the time spent in the master's programme.

[1]California State Polytechnic University, Pomona and Cal-Bridge Program, Pomona, CA, USA. [2]Vanderbilt University and Fisk–Vanderbilt Master's-to-PhD Bridge Program, Nashville, TN, USA. [3]University of Southern California and California Consortium for Inclusive Doctoral Education, Los Angeles, CA, USA. *e-mail: alrudolph@cpp.edu

**Table 1 | A comparison of PhD bridge programmes**

| Programme | Year founded | Student entry point | Number of institutions | Fields | Number of students per year | Total number of students in programme |
|---|---|---|---|---|---|---|
| Fisk–Vanderbilt | 2004 | Master's | 2 | Astronomy, Physics and Material Science, Biology, Chemistry | 10–20 | 146 |
| Columbia | 2008 | Post-baccalaureates | 1 | Multiple science and engineering fields and economics | 8–10 | 65 |
| APS Bridge | 2013 | Master's | 6 out of >25[a] | Physics (including Astronomy in Physics PhD programmes) | 40 | >100 |
| Cal-Bridge | 2014 | Undergraduate | 25 | Physics and Astronomy | 40 | 99 |

[a]Six programmes are APS Bridge Program sites. More than 25 are APS Bridge Program partnership institutions, which do not have the same processes of vetting or expectations of programming as the original bridge sites.

The National Academy of Sciences report '*Expanding Underrepresented Minority Participation: America's Science and Technology Talent at the Crossroads*' highlights two key priorities for broadening participation in the STEM workforce[17]. To address their first priority (undergraduate retention and completion) they propose that higher education institutions provide "strong academic, social and financial support…along with programs that simultaneously integrate academic, social, and professional development." For the second priority (the transition to graduate study) they "encourage programmes that support the transition from undergraduate to graduate education and provide support in graduate programs."

The design and implementation of each of the PhD bridge programmes described here reflect these priorities. Table 1 summarizes the main similarities and differences between the various programmes. Elements that are common to all of the programmes include: no use of the physics GRE in admissions, use of holistic admissions methods and the use of interviews in the selection process.

### Fisk–Vanderbilt Master's-to-PhD Program

The first PhD bridge programme in physics and astronomy was the Fisk–Vanderbilt Master's-to-PhD Program. Founded in 2004, the programme is a partnership between Fisk University (a historically Black college and university) and Vanderbilt University (a research university that is located only two miles away). At the heart of the Fisk–Vanderbilt programme is "the explicit goal…that its students will be well known by the Vanderbilt faculty by the time that they are ready to apply to the Vanderbilt PhD programme of their choice. Indeed, fostering individual mentoring relationships between Fisk students and Vanderbilt faculty is at the very heart of the bridge programme, and is the guiding principle for all other programmatic design considerations"[18].

To that end, the Fisk–Vanderbilt programme has five key elements:

- Students enter the Fisk master's programme, where they are given full financial support throughout the two-year master's phase and the first year of their PhD.

- Each student is jointly mentored by faculty members from both Fisk and Vanderbilt.
- Students meet at least twice per year with the bridge programme Executive Director to review their progress and gain path-planning guidance beyond what is received from their master's committee.
- Students participate in supervised research at Fisk, Vanderbilt or both.
- Students must maintain at least B grades in all graduate courses, including at least one core PhD course at Vanderbilt. Cross-registration privileges have been negotiated between the two schools so that a course taken at either University may count toward both the master's and PhD, mitigating the extra time taken to earn a master's degree before the PhD.

In addition to these key elements of the programme, Fisk–Vanderbilt uses an innovative admissions process that maintains a clear focus on scientific potential, leadership and perseverance. Evidence for these qualities is gleaned from performance in individual courses and/or improvement over time.

The combination of selecting students through this research-based holistic approach with individualized mentoring, exposure to research and financial support has led to the success of the Fisk–Vanderbilt Program. Of the 146 students that have enrolled so far, 58% are Black/African American, 24% are Latinx/Hispanic, 3% are Indigenous/Native American or Pacific Island, and 15% are White or other; 56% percent of the students are women; and over 90% are from traditionally underserved populations (those who are first generation, low-income or have physical or learning disabilities). Since its inception, bridge students have earned over 95 master's degrees and 33 PhDs. The retention to the PhD is 85% and the eight-year PhD completion rate is 89% (well above the national averages). Bridge PhD graduates have been extraordinarily successful in finding STEM employment before graduating: nineteen are employed in academia, including three in tenure-track faculty positions; seven are working in industry; and five are engaged in research at national laboratories. Since 2006, Fisk has been the top producer of Black master's degrees in physics. The Fisk–Vanderbilt Program has also been a model for many of the PhD bridge programmes that followed.

### Columbia Bridge-to-the-PhD Program

Founded in 2008, the Columbia Bridge-to-the-PhD Program takes a slightly different approach. Scholars selected for the programme are not enrolled as master's students, but rather as post-baccalaureates. In this capacity they are provided with an intensive research

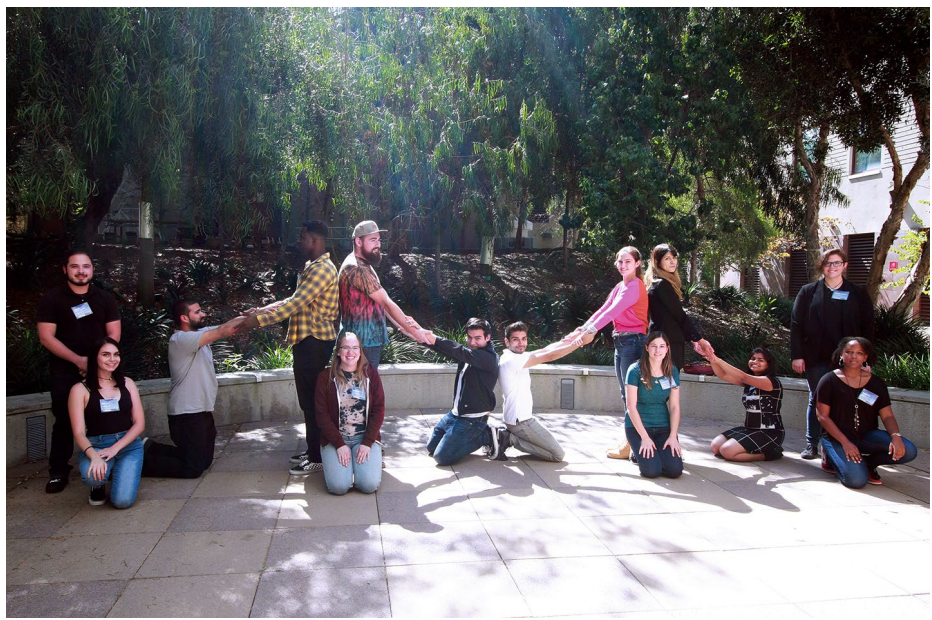

**Fig. 1 | Cal-Bridge South Cohort 5.** Pictured here are fourteen members of Cohort 5 of Cal-Bridge South (the southern California regional programme) attending the fall 2018 orientation. An additional ten scholars were selected as part of Cohort 5 of Cal-Bridge North (the northern California regional programme) that year. Credit: Cal-Bridge Program

experience, coursework and mentoring. Every year, eight to ten new bridge participants are hired as full-time Columbia University research assistants for two years. They conduct research under the mentorship of faculty members, postdoctoral researchers and graduate students, are provided with funds to support professional expenses, and are eligible for university benefits. The participants enrol in at least one course per semester and attend monthly one-on-one progress meetings with the bridge's director. The programme also organizes a number of professional development workshops, access to GRE test preparation and draws on university resources to ensure success while at Columbia and facilitate application to PhD programmes.

Now in its 12th year, the programme supports 16 participants in ten departments, including all branches of science, several engineering departments and economics. Bridge alumni have gone on to PhD programmes across the country and, at the last count, 12 of its alumni have received their PhD, with another half-dozen due to finish in the next year.

### American Physical Society Master's-to-PhD Bridge Program

In 2013 the American Physical Society (APS) founded the national physics Master's-to-PhD Bridge Program[19]. This programme recruits applicants nationally and then shares those applications with over 25 vetted PhD programmes that consider accepting them into a PhD bridge programme in their own department (the one exception to this model is the CSU Long Beach master's programme, which acts more like a traditional stepping-stone degree option described above).

Applications open on 15 April, which is after the deadline for candidates to make PhD decisions. Of the applicants to the APS Program, two-thirds were denied entrance to all of the PhD programmes that they applied to, whereas the other third did not apply at all, often due to concerns that their GRE scores or grades would preclude them from admission. Among the strategies that the programme uses to recruit applicants is to ask undergraduate departments to nominate students that they think would be successful if given the right opportunity and to ask graduate departments to identify applicants that they did not accept and then to encourage those students to apply to the APS Bridge Program.

These recruiting strategies have worked well—there were many more applicants than could be placed at the participating sites—leading to the expansion of the number of sites from the original six to the current >25. The number of placed students has grown from 12 in 2013 to 40 in 2016, with more than 100 students placed in those four short years. Their retention rate is 92%, which compares favourably to the national average in physics of 59% (ref. [20]).

In addition to acting as a national aggregator for applications, the APS Bridge Program provides crucial oversight structures that contribute to the success of the programme. Chief among these is the initial vetting of the programmes that are allowed to accept bridge students. Programmes are required to apply to be a bridge site (or partner site) with a rigorous evaluation process designed to assure that bridge students will be supported once they have matriculated to an institution. The national bridge programme also holds conferences and other events that are designed to build a national cohort and support structure for bridge students.

In 2018, APS partnered with several other disciplinary societies in the physical sciences and researchers from several universities to expand this model to other fields. Through the National Science Foundation (NSF) INCLUDES Alliance, the Inclusive Graduate Education Network (http://igenetwork.org/) aims to both accelerate participation via bridge programmes and embed more holistic admissions and mentoring into the standard practices of PhD programmes across STEM.

### The Cal-Bridge CSU–University of California PhD Bridge Program

Having started in 2014, the Cal-Bridge Program provides a different bridging model from the post-baccalaureate programmes above[11] by serving rising junior undergraduates and providing them with the support structures needed to successfully matriculate to a PhD, especially those at the UC campuses in the Cal-Bridge network. The intent is to thus help students bypass the need to attend a master's or post-baccalaureate programme and proceed directly from their undergraduate institution to a PhD programme.

The programme is a partnership between nine University of California (UC), 16 CSU and over 30 community college campuses in California, with over 200 physics and astronomy faculties from the three systems participating. Scholars are recruited from the CSU and community college campuses in the network, with the help of local faculty and/or staff liaisons at each campus. Community college students transfer to a participating CSU to join the programme.

Following the Fisk–Vanderbilt model, Cal-Bridge uses research-validated selection methods to identify students from underrepresented groups who display strong socioemotional competencies — along with academic potential — and provides them support with four pillars: (1) financial support; (2) intensive, joint mentoring by the CSU and UC faculty members; (3) professional development workshops; and (4) exposure to a wide variety of research opportunities, including at the participating UC campuses. Each of these pillars is an essential support structure for the scholars as they progress towards applying and matriculating to a PhD programme. The innovation of the Cal-Bridge Program is to provide these known, high-impact practices together over an extended two-year period. Furthermore, the programming is provided by faculty members both at the scholars' home institutions and at the institutions (UC campuses) where the scholars hope to matriculate to obtain their PhD.

Like the other bridge programmes described above, Cal-Bridge has had success in its first few years. The programme has grown from a first cohort of five scholars in 2014 to 40 scholars across the state in the sixth cohort selected in fall 2019 (Fig. 1 shows the fifth cohort of the programme). Across six cohorts there have been 99 scholars, of whom 72 are URMs and 44 are women (including 21 women of colour). Of the 33 scholars who have applied to PhDs in the first four years, 27 (82%) are currently enrolled in a programme (including ten in the UC physics and astronomy programmes). Another five are enrolled in master's programmes, hoping to eventually progress to a PhD. Seven have won NSF graduate research fellowships and four more received an honourable mention.

## Lessons learned

Although there are distinct differences among bridge models in both underlying philosophy and practical implementation, some overarching lessons have emerged from 15 years of experience with bridge programmes. We focus here on four that are foundational and common to all programmes.

**Traditional graduate admission metrics miss many talented students.** There is a growing realization in the astronomy community that many traditional admissions criteria effectively exclude qualified applicants on the basis of their socioeconomic and ethnic or racial background, while also doing a poor job at predicting who will succeed once in a PhD programme. The American Astronomical Society (AAS) recently empanelled a task force to survey the research on graduate astronomy admissions and recommended that PhD programmes adopt more holistic admissions practices while carefully studying their own outcomes to learn which criteria really help predict success[21].

Overall, bridge students have been extremely successful in graduate school, demonstrating their ability in PhD-level coursework, publishing in top research journals and earning competitive national fellowships at over twice the rate of students that are on the traditional PhD track. The engagement of faculty members from PhD-granting institutions in mentoring bridge students as well as the subsequent success of those students in graduate school has expanded faculty members' conception of a successful PhD student and promoted change in admissions, inclusion and retention practices at their home institutions. For example, due to the success of bridge students in the programme, the astronomy track at Vanderbilt University has abandoned traditional admissions metrics such as the GRE, as have a number of University of California

PhD programmes. Arguably, bridge-student success has begun to influence a national movement towards holistic admissions at such institutions as the University of Washington; University of Texas, Austin; Harvard University and the University of Arizona.

**An asset- rather than deficit-based perspective promotes success.** Despite the evidence of student talent, bridge programmes have sometimes been viewed — both internally and externally — as a remedial way to 'fix' deficiencies in the student. The physical-science community has constructed its training under the common misperception that scientific talent is an internal spark of genius found in certain people and missing in others. When combined with societal inequities that lead to vast differences in student training, this fixed mindset results in a damaging conclusion that the student, not the system, is deficient. Although the community is beginning to make progress in countering this myth, a student deficit mentality has pervaded the physics and astronomy cultures and can be present even in mentors who seek to broaden participation.

Bridge programmes, on the other hand, can help PhD programmes reframe to asset-based thinking. The holistic criteria used to select bridge scholars typically include assessment of socioemotional skills such as perseverance, creativity, conscientiousness, realistic self-appraisal, leadership and a focus on long-term goals. These skills are often as or more important than academic preparation in navigating academia and supporting a student's success in completing a PhD. The ability to recognize these qualities as assets both diversifies the field and raises the quality of students in PhD programmes. Furthermore, students from bridge programmes typically bring a more diverse set of life experiences than is typical among most PhD programme students, further enhancing the scientific work.

Following the example of bridge programmes, graduate programmes that address student and faculty growth beyond academia continue to underline the value of the scientist as a whole person. Workshops that focus on social, emotional and physical wellbeing foster deeper engagement[22], and seminars on microaggressions, growth mindset and imposter syndrome can help spur cultural awareness among the faculty. Explicit discussions of the unwritten rules of academic culture can help set expectations in the research laboratory and forestall misunderstandings between the student and research advisor. Practical skill-based workshops on, for example, Python, LaTeX or machine learning are also important. In some bridge models these have been taught by senior bridge students to cement their expertise and their standing as an expert in the department.

**An extensive mentoring network is key to ensure student success.** Students in bridge programmes routinely cite mentoring as the most important element of the programme, even above financial aid. Effective mentoring is a key ingredient for success in education at all levels and yet little to no training is provided to budding scientists or faculty members. Because mentoring is a time-intensive activity, there can be a perceived tension in academia between mentoring students and excelling in research and teaching, with the result that tenure and promotion committees have discouraged junior faculties from mentoring bridge students. When not actively discouraged, exceptional mentoring is indifferently rewarded compared with the much more robust rewards afforded to research or teaching.

Recognizing the need to tap into a diverse set of skills and life experiences, many bridge programmes use a mentoring network model to provide a scaffold of support for bridge students. For example, a mentor from the student's home institution may be paired with a mentor from a PhD granting institution[11,16]. This second mentor provides the perspective of faculty from a PhD-granting institution to add to the more intimate knowledge the home institution mentor typically has about the students. In many cases, the mentor on the other side of the bridge plays a valuable role as an advocate for the student during the graduate admissions process.

Nearly all of the bridge programmes also recognize the value of including mentors closer to the career stages of the student in the mentoring network. Near-peer mentors (postdocs or advanced graduate students) or peer mentors (more advanced undergraduates or early stage graduate students) are invaluable as a tangible role-model for the next career stage, and their recent lived experience makes them less intimidating to approach with questions and concerns, to complement more senior mentors who provide their professional experience and wide scientific network. Access to emotional support and a safe space provided by early career mentors can help students deal with the stress of transitioning to a new role and environment. An accountability partner can provide mutual support to meet deadlines and a coach can help to navigate difficult conversations. A key aspect of an effective mentoring model is access to role models of colour and a supportive cohort of other underrepresented minority scholars, as it promotes a sense of belonging and improves performance[23]. Given the small number of physics and astronomy students of colour (especially with intersections of other underrepresented groups), this is a challenge for all bridge programmes as it is for PhD programmes themselves[24]. Mentoring through virtual forums such as VanguardSTEM (a live monthly virtual 'meetup' that supports women of colour; www.vanguardstem.com) has served an important role in connecting marginalized groups in STEM.

**Financial barriers disproportionately exclude URM talent.** Although holistic admissions can widen the pathways to PhD programmes for students from underrepresented groups, financial barriers are a considerable reason students leave a pathway.

The cost of higher education is a barrier to entry that disproportionately affects marginalized groups. At the undergraduate level, the cost of higher education is a considerable barrier to success for many students; for example, despite the 'low cost' of the CSU ($7,000 per year tuition), more than half of CSU students receive state and/or federal aid. The average Cal-Bridge scholarship of $8,000 fills the unmet needs of these students, supplementing the aid that they already receive. Most scholars work 20–30 hours per week in the absence of this aid, impeding their ability to focus on academics.

Even the cost of applying to graduate school — when taking into account standardized tests, test preparation, application fees and travel — can deter talented prospective students. Although students are told that PhD programmes provide financial support throughout graduate education, the reality often differs. Some PhD programmes that are rich in resources can guarantee research support for all five to six years that a typical PhD can take. Others condition research support on having a funded research advisor. Graduate students are often expected to serve as teaching assistants, teaching 20–40 hours a week in addition to doing research. Whether a teaching assistant or on research support, typical graduate school stipends are at a subsistence level at best. Students from lower socioeconomic backgrounds — especially those students who have financial obligations beyond their own, such as helping out their low-income family — are often on the edge of financial insolvency, making them more vulnerable to abandoning the PhD for financial reasons; for example, the cost of transportation to attend class or the need to buy a laptop for work has derailed students in our bridge programmes. A programme that provides a sufficient stipend, tuition and health benefits is critical to student retention, particularly for vulnerable groups.

## Discussion and conclusions

Systemic change is needed to reduce inequities in STEM. Bridge programmes are proving to be an engine of such change by helping facilitate the fraught transition from undergraduate to graduate education. Partnership programmes such as the Fisk–Vanderbilt and Cal-Bridge connect minority-serving institutions that enrol large numbers of Black and Latinx students with PhD-granting universities where these groups are underrepresented. Columbia, Fisk–Vanderbilt and APS all provide a pathway for post-baccalaureate students to successfully progress into a PhD programme, whereas Cal-Bridge cultivates talent before students apply to graduate school, providing research experience as well as support for the transition. The APS Bridge Program model, which is being replicated in new professional societies (including the AAS) through the Inclusive Graduate Education Network, creates a separate application process and national marketplace for students, providing a safety net, of sorts, to ensure that talented students from underrepresented backgrounds do not fall through the cracks of a broken system.

Leaders of all of these bridge programmes recognize that — in a field where participation trend lines for Black, Latinx and Indigenous students have effectively been flat — progress in representation will come one student at a time. We cannot afford to lose interested students who have navigated the educational system to the point where they are on the cusp of its final stage. Furthermore, to sustain gains made through these programmes and create cultures within STEM that are healthier for everyone, the astronomy and physics communities cannot settle for bridges alone. A common critique of these programmes is that they are very expensive relative to the number of students they serve (though much of this cost is due to the escalating costs of higher education). Equity work means not only closing gaps through special programmes but pursuing an iterative system redesign that will help keep them closed.

In this system redesign, bridge programme components provide resources that many disciplines do not have. Basic tenets of bridge programmes' educational models — such as more equity-based, holistic admissions and careful and intensive mentoring — can be applied to graduate programmes as a matter of standard practice. These practices are not new or unique to bridge programmes; rather, as recognized by the recent AAS Task Force on Diversity and Inclusion in Astronomy Graduate Education[21], they are the best available practices for selecting and serving graduate students. After seeing the bridge model in action, Vanderbilt University's Astronomy programme adopted bridge admissions criteria, including structured interviews that are focused on socioemotional competencies. And across the field of astronomy, more and more programmes are adopting equity-based holistic admissions and eliminating GRE score requirements that exclude students and deter some from even applying.

Philosophies of merit, excellence and success affect not only student opportunities at the point of admission, but also how students are educated and mentored. Traditional definitions of merit and success create barriers to underrepresented students who have the ability and drive to attain a PhD but are currently not doing so. Efforts to accelerate participation at the graduate level and beyond require systemic change in these philosophies through transformation in the mindsets of faculty within PhD programmes about what constitutes a successful student and how they can be supported to succeed. The students who progress through these bridge programmes provide powerful evidence to counter and dismantle conventional, racialized mindsets about merit and success, and the principles elucidated by bridge programmes can help guide PhD programmes that say they are committed to equity and inclusion to make good on that commitment.

## Author contributions

## Competing interests

## Additional information